# Magnetic field induced ferroelectric loop in Bi$_{0.75}$Sr$_{0.25}$FeO$_{3-\delta}$

Bohdan Kundys, Antoine Maignan, Christine Martin, Ninh Nguyen, Charles Simon,

Laboratoire CRISMAT, UMR 6508, CNRS  ENSICAEN,

6 Bd. du Maréchal Juin, 14050 Caen Cedex, France

## Abstract

Magnetic field induced ferroelectric hysteresis loop observed in Bi$_{0.75}$Sr$_{0.25}$FeO$_{3\text{-delta}}$ is of prime importance. The coexistence of antiferromagnetism and weak ferromagnetism is responsible for the original magnetoelastic and magnetoferroelectric properties. Upon external magnetic field application, the existence of a magnetostrictive effect supports a structural transition towards a homogeneous antiferromagnetic and ferroelectric phase. The magnetic field induced polarization is among the highest reported for BiFeO$_3$ based systems in either thin film or bulk forms (Pr=96μC/cm$^2$ at 10T) while the ferroelectric coercive field is among the lowest reported (Hc=661(V/cm) at 10T). These properties make this material very attractive for technical applications.

E mail: charles.simon@ensicaen.fr

Recent interest in the magnetoelectric (ME) effects in solids [1-5] has stimulated both research of new materials or reexamination of existing materials. Materials that are electrically polar and anti (or ferro) magnetic, present particular interest from the stand point of the both basic science and (ME) applications. This is basically due to the mutually exclusive nature of magnetism and electric polarization phenomenon in most of solids [6,7]. Materials in which ferroelectricity and anti (or ferro) magnetism coexist are a subclass of materials called multiferroics. Among rare materials combining ferroelectricity and antiferromagnetism, the BiFeO$_3$ (BFO) occupies a unique place as these properties exist at room temperature. Although BFO is known for a long time, it is still a subject of intensive investigations [8-10] aiming at the improvement and understanding of its properties. The stoechiometric bismuth ferrite BiFeO$_3$, crystallizes in a rhombohedrally distorted perovskite structure [11,12]. Its antiferromagnetic and ferroelectric ordering temperatures are $T_N$=643 K [12] and $T_C$=1143 K [13] respectively. Its ordered magnetic structure corresponds to an incommensurate long range cycloidal spiral [13]. The effect of chemical substitutions on the ferroelectric and magnetic properties has been already the focus of numerous studies [14-24]. Among all these reports, the Sr$^{2+}$ substitution for Bi$^{3+}$ [24] leads to coexistence of the both weak ferromagnetic and antiferromagnetic interactions. This has attracted our attention as such coexistence may lead to interesting magnetic field dependent ferroelectric properties. In the present letter, we report on the magnetic field induced appearance of an additional contribution to the conductive dc current that most likely originates from the dipole reorientation in $Bi_{0.75}Sr_{0.25}FeO_{3-\delta}$ ceramics.

The $Bi_{0.75}Sr_{0.25}FeO_3$ polycrystalline sample was prepared in a close vessel by conventional solid state reaction. A mixture of $Bi_2O_3$, $SrO_2$ and $Fe_2O_3$ precursors in the atomic ratios of Bi:Sr:Fe = 0.75:0.25:1 was grinded. The bars obtained by pressing the powder were then put in an alumina crucible which was then inserted in a silica tube. After sealing under primary vacuum, the vessel was heated for 12h at 900°C and then cooled to room temperature at 100°C/h. In contrast to the rhombohedral symmetry found at room temperature in the pure $BiFeO_3$ compound, the X-ray diffraction pattern of the obtained ceramic bars were refined in the cubic space group $Pm3m$ (a = 3.951Å). This transition change from rhombohedral to cubic induced by the $Sr^{2+}$ for $Bi^{3+}$ substitution is consistent with the previously reported study [33]. In order to check for oxygen nonstoechiometry, a Mössbauer study was also performed. When compared to the results obtained for a rhombohedral pristine $BiFeO_3$, prepared with the same process, an additional iron site (~ 7 %) is observed with Mössbauer parameters typical of coordination lower than six which is attributed to the presence of tetragonal pyramids $FeO_5$. This indicates the existence of 3.5 % vacancy, i.e. an average iron state close to ~ 3.04. These data, also confirmed by chemical titration, strongly support that the present heterovalent substitution on the Bi site does not create $Fe^{4+}$ but is rather compensated by oxygen vacancies. Ferroelectric and magnetization measurements were carried out in a PPMS Quantum Design cryostat using a Keithley 6517A electrometer for quasi-static voltage-current measurements. The magnetic field was applied perpendicular to the direction of the electric one. The electrical contacts to the sample were made using a conductive silver paste. The magnetostriction was measured (from 6T to -6T) using a three terminal capacitance dilatometer technique [30] in a PPMS cryostat.

The coexistence of antiferromagnetic weak ferromagnetic interactions simultaneously in our $Bi_{0.75}Sr_{0.25}FeO_{3-\delta}$ sample (Fig.1) is consistent with the previous published data for the $Bi_{1-x}Sr_xFeO_3$ with $x$ = 0.2, 0.4, and 0.6 [25]. A weak ferromagnetic component is seen in the low magnetic field region. An irreversible magnetization is observed in this area. The weak ferromagnetic order itself can be understood as a result of non-collinear (canted) spin arrangements in two sublattices [25]. As the magnetic field increases, the ferromagnetic component saturates at about 1T and the antiferromagnetic component dominates.

To determine how this magnetic behavior influences the electric properties of the sample, I-V characteristics were measured at 270K within different applied magnetic fields. This temperature was chosen to minimize the dielectric losses. The sample was initially cooled from 320K down to 270K in electric field of 570V/cm in zero magnetic field. Then successive I-V curves were recorded at magnetic fields of 0, 3, 7, and 10 T respectively.

While a very flat I-V curve is observed at zero magnetic field, the increase of magnetic field generates a spectacular difference in the I-V curves (figure 2). In particular a growing peak as the magnetic field increases is induced. Moreover, the position of the peak shifts towards lower electric fields as the magnetic field increases. The results obtained from the current integration reveal the ferroelectric- hysteresis loops P(E) shown in Fig.3. In addition to the ferroelectric hysteresis loops, the magnitude of the magnetic field has two effects on the loops: it makes increasing the spontaneous polarization $(P_s)$ and significantly decreasing coercive field $(H_c)$ (Fig.3.insets). This polarization is among the highest reported for $BiFeO_3$ based systems in either thin film or bulk forms ($P_r=96\mu C/cm^2$ at 10T) while the ferroelectric coercive field is among the lowest reported ($H_c=661(V/cm)$ at 10T). These properties make this material very attractive for technical applications.

As observed in Fig. 3, the P(E) slope is rather identical near P=0. This implies that the dielectric permittivity is independent on the magnetic field. Furthermore, no magneto-capacitance is detectable in our experiment suggesting that it is not crucial to have magnetocapacitance effect to observe a magnetic field induced ferroelectric hysteresis loop. It also has to be noted that magneto-capacitance effect may have different contributions that not necessarily reflect the ME coupling [26-28]. As the sample in not ferroelectric in absence of applied magnetic field, one may assume that it should undergo a structural distortion in which electrical dipoles appear. In order to verify this assumption, we have measured magnetostriction of the sample. To keep the same configuration of experiment the magnetic field was applied perpendicularly to the direction in which the change of dimension was measured. As one can see, the sample undergoes maximum of structural deformation approximately at the same magnetic field region where the step-like slope in the magnetization loop is observed (Fig.4). One may suggest that the structural distortion is a result of AFM spin alignment due to the exchange interaction dependence on the distance between the Fe ions. Indeed, application of the high magnetic field was reported to induce a phase transition from the spatially modulated antiferromagnetic spin structure to a homogeneous one in a pure $BiFeO_3$ [29].

In conclusion, we have performed a study on magnetic, magnetoelastic, and electric properties of $Bi_{0.75}Sr_{0.25}FeO_{3-\delta}$. The nonstoichiometry induced by 25% Sr doping results in a presence of both weak ferromagnetic and antiferromagnetic components and manifests itself in the interesting magnetic field induced electric properties. As the magnetic field is increased, the weak ferromagnetism is saturated and overwhelmed by the antiferromagnetic ones that reflect itself in the shape of the magnetization loop and lead to the structure deformation accompanied by an appearance of the displacement current. The I-V characteristics depend strongly on the magnetic field and show presence of a ferroelectric contribution for H>3T. We suggest that no magneto-capacitance effect is crucial to observe magnetic field induced ferroelectric hysteresis loops. In the present oxygen deficient $Bi_{0.75}Sr_{0.25}FeO_{3-\delta}$ compound, the local disorder linked to oxygen vacancies is merely responsible for the inhomogeneous antiferromagnetic state. Thus, it is assumed that magnetic field induced long range homogeneous antiferromagnetic spin structure is favorable to observe ferroelectricity in non collinear antiferromagnets. These results provide an interesting approach towards the magnetically modulated ferroelectric devices.

**Figure:**

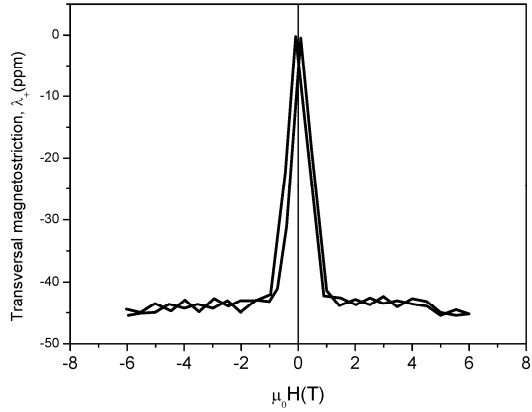

Fig.4. Transverse magnetostriction ($\Delta L/L$) of Bi$_{0.75}$Sr$_{0.25}$FeO$_{3-\delta}$ at 270K

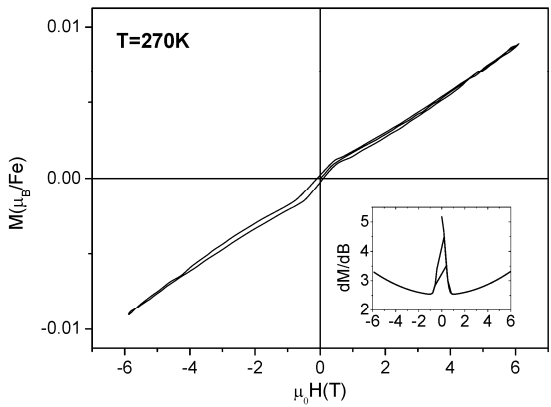

Fig.1. Magnetization loop: at low magnetic field, there is a small ferromagnetic component coexisting with an antiferromagnetic phase which gives the large susceptibility. In the inset, the susceptibility dM/dH.

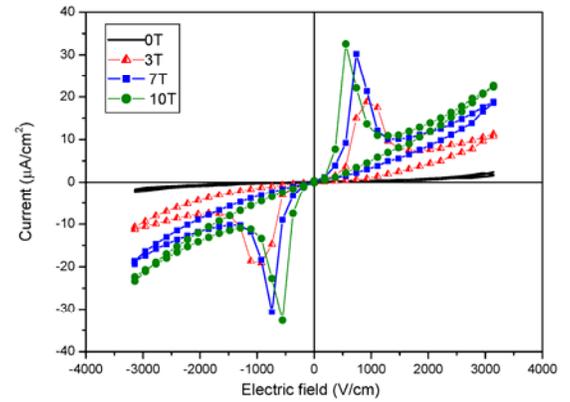

Fig.2. Voltage – current characteristics taken at 270K at different magnetic fields. The raw data reveal magnetic field induced additional peaks on a background due to leakage current.

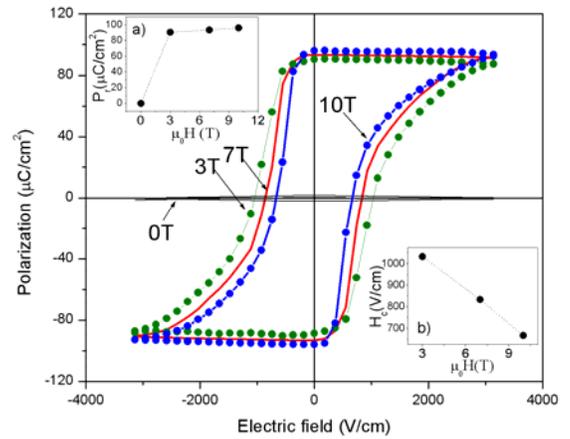

Fig.3. Ferroelectric loops obtained as a result of current integration at 270K (after application of the leakage minimization procedure). Inset: a) Magnetic field dependence of remnant polarization $(P_r)$ and b) ferroelectric coercive field $(H_c)$ respectively.